\documentclass{wnna}
\usepackage{natbib}\usepackage{txfonts}\usepackage{balance}
\usepackage{graphicx}
\usepackage[a4paper]{hyperref}
\idline {1}{101}

\def\ltsim{\footnotesize{\mathop{\raisebox{-.4ex}{\rlap{$\sim$}}
\raisebox{.4ex}{$<$}}}}

\begin{document}

\title{Sensitivity of the NEMO detector to galactic microquasars}

\author{
C. \,Distefano \inst{1} for the NEMO Collaboration}

\authoremail{distefano\_c@lns.infn.it}

\institute{Istituto Nazionale di Fisica Nucleare -- Laboratori
Nazionali del Sud, Via S. Sofia 62, I-95123 Catania, Italy}

\authorrunning{Distefano C.}
\titlerunning{Sensitivity of the NEMO detector to galactic microquasars}

\abstract{We present the results of Monte Carlo simulation studies of the capability of the proposed NEMO km$^3$ telescope
to detect TeV muon neutrinos from Galactic microquasars. In particular we determined the detector sensitivity to each known
microquasar, optimizing the event selection in order to reject the atmospheric background. We also determined the expected
number of source and background events surviving the selection.

\keywords{Microquasars --- Neutrino telescopes --- NEMO}
}


\maketitle{
}

\section{Introduction}

Microquasars are Galactic X-ray binary systems which exhibit
relativistic jets, observed in the radio band \citep{Chaty05}.
Several authors propose microquasar jets as sites of acceleration of charged particles
up to energies of about $10^{16}$ eV, and of high energy neutrino production.
According to present models, neutrinos could be produced both in p$\gamma$ \citep{Levinson01,Distefano02}
and pp interaction scenarions \citep{Bednarek05,Aharonian05,Christiansen05,Romero05}.

The aim of this paper is to study the possibility to detect neutrinos from
known microquasars with the proposed NEMO-km$^3$ telescope \citep{Migneco06}.
In particular, for each microquasar we calculated the expected sensitivity,
optimizing the event selection to reject the atmospheric neutrino and muon backgrounds.

We also calculated, according to the present theoretical models, the expected number of microquasar events 
that survive the event selection. We compared the expected source signal with the remaining background to
establish its detectability.

The NEMO proposal is made in the context of th KM3 prpject for a km$^3$ detector in the Mediterranean sea \citep{km3net}.

\section{The NEMO km$^3$ detector}

The NEMO-km$^3$ telescope, simulated in this work, is a square array of
$9\times9$ towers with a distance between towers of 140 m.
In this configuration each tower hosts 72 PMTs (with a diameter of 10"), namely 5832 PMTs for the
whole detector with a total geometrical volume of $\sim0.9$ km$^3$.
We considered an 18 storey tower; each storey
is made of a 20 m long beam structure hosting two optical modules (one downlooking and one looking horizontally) at each
end (4 OMs per storey). The vertical distance between storeys is 40 m. A spacing of 150 m is added at the base of the
tower, between the anchor and the lowermost storey.

The detector response is simulated using the simulation codes developed by the ANTARES
Collaboration \citep{Amram02, Becherini06}, modified for a km$^3$ telescope \citep{Aiello07}. In the simulation codes, the light
absorption length, measured in the site of Capo Passero
($L_a\approx 68$ m at 440 nm \citep{Riccobene07}), is taken into account. Once
the sample of PMT hits is generated, spurious PMT hits,
due to the underwater optical noise ($^{40}$K decay), are
introduced, with a rate of 30 kHz for 10"
PMTs, corresponding to the average value measured in the Capo
Passero site.

\section{Calculation of the sensitivity}

The detector sensitivity was calculated according to the \citet{Feldman98}
approach. The 90\% c.l. sensitivity to a neutrino flux coming from a microquasar
is given by
\begin{equation}
f_{\nu,90}=\frac{\overline{\mu}_{90}(b)}{N_\mu^m}f_{\nu}^{th},
\end{equation}
where $\overline{\mu}_{90}(b)$ is the 90\% c.l. average upper
limit for an expected background (atmospheric neutrinos + muons)
with known mean value $b$ and no true signal \citep{Feldman98},
$f_{\nu}^{th}$ is the theoretical neutrino energy flux from a given microquasar that induces
a mean signal $N_\mu^m$. During the calculation, an event selection is applied
in order to optimize the sensitivity, as described in \citet{Aiello07}.
Detailed calculation of the sensitivity for the proposed NEMO
km$^3$ telescope to a generic point-like muon neutrino source
are presented in \citet{barcellona}.

\begin{figure}[h]
\resizebox{\hsize}{!}{\includegraphics[clip=true]{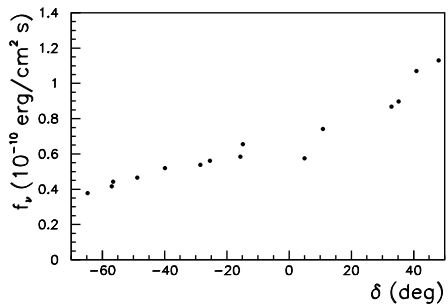}}
\caption{ \footnotesize NEMO-km$^3$ sensitivity to neutrinos from microquasars versus source declination,
for a livetime of 1 year. The worsening of the sensitivity with increasing declination is
due to the decrease of the source visibility.}
\label{fig:sensi-ene-flusso-delta-1years}
\end{figure}

The detector sensitivity was calculated for a livetime of 1 year,
simulating a neutrino flux with spectral index $\Gamma=2$ in the
energy range 1 - 100 TeV. The study was carried out for each
microquasar, since the sensitivity is a function of the source
astronomical declination. Results are given in
Tab. \ref{tab:sensitivity} $f_{\nu,90}$. Fig.
\ref{fig:sensi-ene-flusso-delta-1years} shows the detector
sensitivity for the studied microquasars as a function of the
declination; the sensitivity flux limit increases with increasing
declination, due to the decrease of the time per day spent by the
source below the Astronomical Horizon (with respect to the
latitude of the Capo Passero site). In Tab. \ref{tab:sensitivity},
we present only results concerning microquasars which may be
observed by a telescope located in the Capo Passero site (i.e.
with declination $\delta\ltsim+54^\circ$).

\begin{table}[h]
\caption{Detector sensitivity to neutrinos from
microquasars: the sensitivity $f_{\nu,90}$ (expressed in erg/cm$^2$ s) is calculated for an
$\varepsilon_\nu^{-2}$ neutrino spectrum in the energy range 1 -
100 TeV, for a detector livetime of 1 year. The corresponding
values of the search bin angular radius $r_{bin}$ and
the source declination $\delta$ are also given, both expressed in degrees.} \label{tab:sensitivity}
\begin{center}
\begin{tabular}{lccc}
\hline
\\
     {\bf Source name}    & $r_{bin}$  & $f_{\nu,90}$   &  $\delta$  \\
\hline \hline
{\it Steady Sources} \\
\hline
  LS 5039                           & 0.9     & $6.5\cdot10^{-11}$    & -14.85    \\
  Scorpius X-1                      & 0.7     & $5.8\cdot10^{-11}$    & -15.64    \\
  SS433                             & 0.8     & $5.7\cdot10^{-11}$    & ~~4.98    \\
  GX 339-4                          & 0.5     & $4.7\cdot10^{-11}$    & -48.79    \\
  Cygnus X-1                        & 0.7     & $9.0\cdot10^{-11}$    & ~35.20    \\
\hline
{\it Bursting Sources} \\
\hline
  XTE J1748-288                     & 0.9     & $5.4\cdot10^{-11}$    & -28.47    \\
  Cygnus X-3                        & 0.8     & $1.1\cdot10^{-10}$    & ~40.95    \\
  GRO J1655-40                      & 0.7     & $5.2\cdot10^{-11}$    & -39.85    \\
  GRS 1915+105                      & 0.8     & $7.4\cdot10^{-11}$    & ~10.86    \\
  Circinus X-1                      & 0.9     & $4.2\cdot10^{-11}$    & -56.99    \\
  XTE J1550-564                     & 0.9     & $4.4\cdot10^{-11}$    & -56.48    \\
  V4641 Sgr                         & 0.9     & $5.6\cdot10^{-11}$    & -25.43    \\
  GS 1354-64                        & 1.0     & $3.8\cdot10^{-11}$    & -64.73    \\
  GRO J0422+32                      & 0.8     & $8.7\cdot10^{-11}$    & ~32.91    \\
  XTE J1118+480                     & 0.7     & $1.1\cdot10^{-10}$    & ~48.05    \\
\\
\hline
\end{tabular}
\end{center}
\end{table}

\section{Expected number of microquasar events}

In Tab. \ref{tab:N_mu_simulated} are given the number of selected
neutrino events from each microquasar, applying the event selection that optimize
the sensitivity in Tab. \ref{tab:sensitivity} and according to the neutrino fluxes
given by \citet{Distefano02}. A detailed analysis considering other
neutrino production models in microquasars is reported in \citet{Aiello07}.
The results in Tab. \ref{tab:N_mu_simulated} refer to an
integration time $\Delta t$ equal to the duration of the
considered burst for the transient sources and to 1 year for the
steady sources. In the same table is given the background
(atmospheric neutrinos + muons) in 1 year of data taking. In this
analysis, it is assumed that transient sources cause one burst per
year, i.e. the number of source events produced in the interval
$\Delta t$ is relative to 1 year observation time.

\begin{table*}
\caption{Expected number of neutrino induced muons from the
microquasar model proposed by \citet{Levinson01}: N$_\mu^{m}$ is the number
of selected muons from each microquasar expected from the
theoretical neutrino energy flux $f_\nu^{th}$ quoted by \citet{Distefano02}, during the time interval $\Delta t$. We
also report the expected number of atmospheric background events
$b$ surviving the event selection and expected in 1 year of data
taking.} \label{tab:N_mu_simulated}
\begin{center}
\begin{tabular}{lcc|cc}
\hline
\\
     {\bf Source name}       & {\bf ${\bf \Delta t}$ (days)} & {\bf ${\bf f_\nu^{th}}$ (erg/cm$^2$ s)} & {\bf ${\bf N_\mu^{m}}$} &  {\bf ${\bf b}$}  \\
\hline \hline
{\it Steady Sources} \\
\hline
  LS 5039                 & 365            & 1.69$\cdot10^{-12}$                        &   0.1             &   0.1         \\
  Scorpius X-1            & 365            & 6.48$\cdot10^{-12}$                        &   0.2             &   0.1         \\
  SS433                   & 365            & 1.72$\cdot10^{-9~}$                        &   76.0            &   0.1         \\
  GX 339-4                & 365            & 1.26$\cdot10^{-9~}$                        &   68.0            &   0.1         \\
  Cygnus X-1              & 365            & 1.88$\cdot10^{-11}$                        &   0.5             &   0.1     \\
\hline
{\it Bursting Sources} \\
\hline
  XTE J1748-288           & 20             & 3.07$\cdot10^{-10}$                        &   0.8             &   0.3         \\
  Cygnus X-3              & 3              & 4.02$\cdot10^{-9~}$                        &   0.8             &   0.1     \\
  GRO J1655-40            & 6              & 7.37$\cdot10^{-10}$                        &   0.6             &   0.1         \\
  GRS 1915+105            & 6              & 2.10$\cdot10^{-10}$                        &   0.1             &   $<0.1$      \\
  Circinus X-1            & 4              & 1.22$\cdot10^{-10}$                        &   0.1             &   0.1         \\
  XTE J1550-564           & 5              & 2.00$\cdot10^{-11}$                        &   $<0.1$          &   $<0.1$      \\
  V4641 Sgr               & 0.3            & 2.25$\cdot10^{-10}\div3.25\cdot10^{-8}$    &   $<0.1\div$1.4       &   0.1         \\
  GS 1354-64              & 2.8            & 1.88$\cdot10^{-11}$                        &   $<0.1$          &   0.1         \\
  GRO J0422+32            & 1$\div$20      & 2.51$\cdot10^{-10}$                        &   $<0.1\div$0.4       &   0.1         \\
  XTE J1118+480           & 30$\div$150    & 5.02$\cdot10^{-10}$                        &   1.0$\div$4.8        &   0.2         \\
\\
\hline
\end{tabular}
\end{center}
\end{table*}

In order to estimate the event rates for non-persistent sources it
is crucial to know their duty cycle. Some of these sources have a
periodic bursting activity: Circinus X-1 has a period of 16.59
days \citet{Preston83}. This means therefore that we expect about
1.5 events per year. Other transient sources show a stochastic
bursting activity. For such cases it is difficult to give an
estimate of the expected event rate. For example, during 1994 GRO
J1655-40 had three radio flares, each lasting 6 days
\citep{Hjellming95}; during the same year GRS 1915+105 emitted 4
bursts \citep{Rodriguez99}. A recent study \citep{Nipoti05} has
shown that microquasars GRS 1915+105, Cygnus X-3 and Scorpius X-1
are in flaring mode 21, 10 and 3 percent of the time,
respectively. The possibility to integrate over more then one
burst could therefore help to detect neutrinos from microquasars.

The search for neutrino events in coincidence with microquasar
radio outbursts could be a tool to reject atmospheric background,
restricting the analysis period to the flare duration $\Delta t$.
Such an analysis technique, already used by AMANDA
\citet{Amanda05}, can improve the detector sensitivity to neutrinos
from transient sources. Referring to the bursts considered in Tab.
\ref{tab:N_mu_simulated} and integrating over the time interval
$\Delta t$ of the bursts, we expect an average background of about
$10^{-3}$ events (muons) per burst. Summing on all the bursting
sources, in Tab. \ref{tab:N_mu_simulated} we count $\sim0.04$
background events, which requires about 5 source events for a
5$\sigma$ level detection with a 70\% probability \citep{IceCube}.
Tab. \ref{tab:N_mu_simulated} shows that we expect $3.4\div9.0$
events in the case of a burst from each of the bursting
microquasars. Therefore, a cumulative analysis could provide a
possible detection of microquasar neutrinos.

\section{Conclusions}

The possibility to detect TeV neutrinos
from Galactic microquasars with the proposed NEMO-km$^3$ underwater
\v{C}erenkov neutrino telescope has been investigated.
A Monte Carlo was carried out to simulate the expected
neutrino-induced muon fluxes produced by microquasars and by
atmospheric neutrinos. The expected atmospheric muon background
was also simulated.
We computed the detector sensitivity for each microquasar,
optimizing the event selection in order to reject the background.
Finally, we applied the event selection and calculated the number
of surviving events.
Our results show that,
assuming reasonable scenarios for TeV neutrino production,
the proposed NEMO telescope could identify microquasars in a few years of data taking,
with a discovery potential for at least few cases above the 5$\sigma$ level,
or strongly constrain the neutrino production models and the source parameters.


\bibliographystyle{aa}

\end{document}